# Design and Implementation of Code Completion System Based on LLM and CodeBERT Hybrid Subsystem

Bingbing Zhang[1], Ziyu Lin[2], Yingxin Su[3]

[1] *Xiamen Institute of Technology, Xiamen, China*

[2] *Google LLC, Seattle, Washington, USA*

[3] *University of California, Davis, USA*

[1]zhangbingbing@xit.edu.cn

[2]zil332@ucsd.edu

[3]cyxsu@ucdavis.edu

**Abstract.** In the rapidly evolving industry of software development, coding efficiency and accuracy play significant roles in delivering high-quality software. Various code suggestion and completion tools, such as CodeBERT from Microsoft and GPT-3.5 from OpenAI, have been developed using deep learning techniques and integrated into IDEs to assist software engineers' development. Researches have shown that CodeBERT has outstanding performance in code summarization and capturing code semantics, while GPT-3.5 demonstrated its adept capability at code generation. This study focuses on implementing a hybrid model that integrates CodeBERT and GPT-3.5 models to accomplish code suggestion and autocomplete tasks, leveraging the context-aware effectiveness of CodeBERT and taking advantage of advanced code generation abilities of GPT-3.5. Evaluated in three main metrics: accuracy, quality of generated code and performance efficiency with various software and hardware, the hybrid model outperforms benchmarks, demonstrating its feasibility and effectiveness. Robustness testing further confirms the reliability and stability of the hybrid model. This study not only further emphasizes the importance of deep learning in the software development industry, but also reveals the potential of synthesizing complementary deep learning models to fully exploit strengths of each model.

- **Introduction**

In recent years, the rapid advancement of software development has led to an increasing demand for efficient coding practices. As programming languages and development tools evolve, developers face the challenge of managing complex codebases and ensuring high-quality software delivery. Code completion technology has emerged as a crucial tool to enhance developer productivity by providing intelligent suggestions and automating repetitive coding tasks. This technology not only reduces the time spent on coding but also minimizes the likelihood of errors, thereby improving overall software quality [1].

The integration of deep learning techniques into code completion systems has garnered significant academic attention. Traditional methods often rely on static analysis and pattern matching, which may not effectively capture the nuances of modern programming languages. In contrast, deep learning-based approaches leverage large datasets and advanced neural network architectures to understand code semantics and context. Among these, CodeBERT has shown remarkable performance in code understanding tasks, while models like GPT-3.5 have demonstrated powerful capabilities in natural language generation [2].

This study aims to explore the potential of a hybrid model that combines the strengths of

CodeBERT and GPT-3.5 for code completion tasks. By integrating the contextual understanding of CodeBERT with the generative capabilities of GPT-3.5, the proposed model seeks to enhance the accuracy and quality of code suggestions. The research will also evaluate the model's performance across various dimensions, including accuracy, generation quality, efficiency, and robustness, using a comprehensive dataset derived from Microsoft's CodeXGLUE benchmark.

- **Literature Review**

Integrating deep learning models has become mainstream in software development to generate more efficient and reliable code. Feng X. et al. [3] introduced CodeBERT, the first bimodal pre-trained model for programming languages (PL) and natural language (NL). Trained on NL-PL pairs and unimodal data, CodeBERT demonstrated strong semantic understanding in code search and code-to-documentation tasks. Pan C. et al. [4] further confirmed its effectiveness in defect prediction with sentence-based and keyword-based extensions (CodeBERT-PS, CodeBERT-PK). Although CodeBERT does not employ abstract syntax trees (ASTs), they suggested integrating AST-based representations could enhance performance.

Wang Y. et al. [5] emphasized the role of ASTs in code completion, noting traditional AST methods lack capacity to capture complete structural patterns. They proposed CCAG, which models flattened ASTs as graphs and introduces an AST Graph Attention Block (ASTGab) with three attention layers to capture dependencies among AST nodes. Subtasks are balanced using uncertainty, and extensive experiments validated its effectiveness in code completion.

Alongside CodeBERT, GPT-3 is another widely recognized model. Brown TB et al. [6] evaluated GPT-3 on multiple NLP tasks in zero-, one-, and few-shot settings, demonstrating strong transfer learning. Despite not being fine-tuned, GPT-3 often matched or surpassed fine-tuned models, with performance improving as model size increased. Its adaptability and text generation ability position it as a competitive candidate for building general-purpose language systems, despite certain limitations.

To support evaluation, Lu S. et al. [7] introduced CodeXGLUE, a benchmark dataset suite covering tasks like code completion and defect detection. It includes baseline systems—BERT-style, GPT-style, and Encoder-Decoder frameworks—providing comprehensive foundations for model comparison. CodeXGLUE enables researchers to assess not only individual task performance but also holistic model capabilities.

- **Experimental Result**

*3.1 Data Introduction*

The dataset used in this study is derived from Microsoft's open source CodeXGLUE benchmark dataset, which is dedicated to the evaluation of code comprehension and generation tasks. In this study, the focus is on using the Python code dataset related to the Code Completion task therein, which has been carefully filtered and preprocessed to ensure data quality and diversity.

*3.2 Model Introduction*

In terms of model architecture, this study proposes a hybrid model architecture based on CodeBERT and GPT-3.5. Among them, CodeBERT is responsible for handling the contextual encoding of the code and extracting the semantic features of the code through a multi-layer transformer structure, while GPT-3.5 serves as a back-end generative model that is responsible for generating high-quality code-completion results based on the contextual features. A feature fusion mechanism is designed between the two models to ensure that the model can fully utilize the advantages of the two different architectures.

Specifically, the CodeBERT model adopts a pre-training-fine-tuning paradigm, where pre-training on a large-scale code corpus is followed by specific fine-tuning for the code-completion task. The model contains a 12-layer transformer encoder with a hidden layer dimension of 768 and an attention head count of 12. GPT-3.5 is used as the generative model, and autoregressive code generation is employed to ensure that the generated code conforms to syntactic specifications and semantic coherence.

The core computation of the following attention mechanism, where Q, K, and V denote the Query, Key, and Value vectors, respectively, and d denotes the vector dimension. By scaling the dot product attention computation, the model is able to effectively capture long distance dependencies in code sequences.

In order to optimize the model performance, this study designs a feature fusion layer that organically combines the contextual features extracted by CodeBERT with the generative capabilities of GPT-3.5. Meanwhile, a dynamic attention mechanism is implemented so that the model can adaptively adjust the feature weights according to different code contexts, thus improving the accuracy and practicality of code completion.

Where represents the feature vector extracted by CodeBERT, denotes the feature vector generated by GPT, and α is a learnable fusion weight parameter (0≤α≤1). By dynamically adjusting the value of α, the model can adaptively balance the importance of the two features according to different inputs.

During the training process, a phased training strategy was adopted, first fine-tuning CodeBERT, then training the feature fusion layer, and finally optimizing the model parameters as a whole. This training approach ensures that each component of the model can give full play to its performance, ultimately forming an efficient code-completion system.

The model training process is optimized using a cross-entropy loss function, where denotes the true label and denotes the model prediction probability. In the prediction stage, the model generates the next code token by conditional probability , where denotes the predicted token at the current moment, } denotes the sequence of all previously generated tokens, is the hidden state, and W and b are the weight matrix and the bias term, respectively.

3.3 Model evaluation

In terms of model evaluation, this study adopts a multi-dimensional evaluation index system to comprehensively measure the performance of the code-completion system. The main evaluation index is Accuracy, which is calculated by the formula:

Also, BLEU scoring is introduced in this paper to assess the quality of the generated code:

Where BP is the penalty factor, is the n-gram weight, and is the n-gram precision.

To evaluate the real-time performance of the model, the average response time (ART) metric is introduced:

Where and denote the code-completion start and end times, respectively, and N is the total number of test samples. These evaluation metrics comprehensively assess the model performance from three dimensions: accuracy, generation quality and efficiency, and provide a reliable quantitative basis for the optimization of the system. Meanwhile, the effectiveness of the proposed method is further verified through comparison experiments with the benchmark model.

- **Experimental environment**

The experiments are conducted in the above hardware and software environments, and all experiments use the same configurations to ensure comparable and reproducible results.The GPU uses NVIDIA A40 to provide sufficient computational power to ensure the efficiency of large-scale model training. Stable versions of deep learning frameworks and related dependent libraries were selected for the software environment to ensure the reliability of the experiments.

4.1 Experimental Analysis

Tables 1 to 4 provide a comprehensive evaluation of the performance of different models, including the baseline model, CodeBERT, GPT-3.5, and the hybrid model, across various dimensions such as accuracy, generation quality, performance efficiency, and robustness. These tables support the analysis of the hybrid model's superiority in achieving higher accuracy, better code generation quality, improved performance efficiency, and strong robustness. Additionally, Figure 1 visually compares the accuracy metrics across models, highlighting the hybrid model's significant improvements in precision, recall, and F1-Score. Figure 2 further illustrates the generation quality metrics comparison, emphasizing the hybrid model's superior performance in BLEU score, code executability, and semantic consistency. Together, these visual and tabular results demonstrate the effectiveness of the proposed hybrid model architecture.

**Table 1**: Accuracy Evaluation

| Model | Accuracy | Precision | Recall | F1-Score |
|---|---|---|---|---|
| Baseline | 0.82 | 0.79 | 0.81 | 0.8 |
| CodeBERT | 0.87 | 0.85 | 0.86 | 0.85 |
| GPT-3.5 | 0.89 | 0.88 | 0.87 | 0.87 |
| Hybrid Model | 0.93 | 0.91 | 0.92 | 0.91 |

**Figure 1.** Accuracy Metrics Comparison Across Models

The experimental results reveal significant advantages of the hybrid model across multiple evaluation metrics. The model demonstrates remarkable accuracy improvements, achieving an F1-Score of 0.91, which represents a substantial 13.75% enhancement compared to the baseline model. This improvement indicates the effectiveness of combining CodeBERT and GPT-3.5 architectures in understanding and generating code sequences.

**Table 2:** Generation Quality Analysis

| Model | BLEU | Code Executability | Semantic Consistency |
|---|---|---|---|
| Baseline | 0.65 | 0.78 | 0.72 |
| CodeBERT | 0.72 | 0.85 | 0.79 |
| GPT-3.5 | 0.76 | 0.87 | 0.83 |
| Hybrid Model | 0.81 | 0.92 | 0.88 |

**Figure 2** Generation Quality Metrics Comparison

Regarding code quality, Figure 2 and Table 3 together illustrate the benefits of integrating strengths of CodeBERT and GPT-3. Code quality is evaluated in three dimensions: BLEU score, Code Executability and Semantic Consistency. BLEU score is a key metric for assessing the quality of machine-translated texts. A higher score indicates the machine-generated texts have a greater similarity to the reference translations.

**Table 3: Performance Efficiency Analysis**

| Model | Average Response Time(ms) | Memory Usage(GB) | Inference Speed(tokens/s) |
|---|---|---|---|
| Baseline | 85 | 4.2 | 156 |
| CodeBERT | 78 | 5.1 | 182 |
| GPT-3.5 | 72 | 5.8 | 195 |
| Hybrid Model | 68 | 6.2 | 213 |

Performance efficiency analysis shows that the hybrid model maintains excellent real-time capabilities, with an average response time of 68ms and an inference speed of 213 tokens per second. These performance metrics indicate that the integration of multiple model components does not compromise computational efficiency, making it suitable for practical development environments where rapid response times are crucial.

Table 4: Model Robustness Analysis

| Test Scenario | Accuracy | Recovery Ability | Stability Index |
|---|---|---|---|
| Normal Input | 0.93 | 0.95 | 0.94 |
| Noisy Input | 0.87 | 0.89 | 0.88 |
| Incomplete Input | 0.85 | 0.88 | 0.86 |
| Abnormal Input | 0.82 | 0.84 | 0.83 |

Furthermore, robustness testing reveals the model's consistent performance across various input scenarios, demonstrating strong generalization capabilities and operational stability. This robust performance across different testing conditions confirms the model's reliability and practical value in diverse programming contexts. The comprehensive results validate the effectiveness of the proposed hybrid architecture, showcasing significant improvements in both accuracy and quality while maintaining optimal efficiency and robustness levels.

In summary, the Hybrid Model emerges as the most efficient in terms of response time and inference speed but at the cost of increased memory usage, reflecting a balance between computational performance and resource demand.

- **Conclusion**

This study investigates the potential of implementing a hybrid model which leverages outstanding context-aware capabilities of CodeBERT and remarkable performance in code generation of GPT-3.5. The multidimensional evaluation index system provides a comprehensive assessment of the hybrid model across code generation accuracy, quality and efficiency aspects. Results validate that the proposed hybrid model surpasses the current benchmark models in all three aspects, offering a promising solution for code completion tasks across various software and hardware environments without comprising robustness. To address the complexities of real-world software development, the hybrid architecture undergoes rigorous testing across different input scenarios to ensure its stability and generality. This study reinforces the significance of deep learning in the software development industry and successfully demonstrates the benefits and feasibility of synthesizing deep learning models. Furthermore, this research paves a path for future studies that explores model fusion to improve software development efficiency and quality.

- **References**
- Alenezi M, Akour M. Ai-driven innovations in software engineering: a review of current practices and future directions[J]. Applied Sciences, 2025, 15(3): 1344.
- Zhou X, Han D G, Lo D. Assessing generalizability of codebert[C]//2021 IEEE International Conference on Software Maintenance and Evolution (ICSME). IEEE, 2021: 425-436.
- Feng Z, Guo D, Tang D, et al. CodeBERT: A pre-trained model for programming and natural languages[C]//Findings of EMNLP 2020. 2020: 1536-1547.
- Liu X, Zhou H, Zhou W, et al. On the versatility of large language models for software engineering tasks[J]. IEEE Transactions on Software Engineering, 2023.
- Brown T B, Mann B, Ryder N, et al. Language models are few-shot learners[J]. Advances in neural information processing systems, 2020, 33: 1877-1901.
- Wei J, Goyal M, Durrett G, et al. Code completion by modeling flattened abstract syntax trees as graphs[C]//Proceedings of the AAAI Conference on

Artificial Intelligence. 2020, 34(05): 9159-9166.
- Lu S, Guo D, Ren S, et al. CodeXGLUE: A Machine Learning Benchmark Dataset for Code Understanding and Generation[J]. arXiv preprint arXiv:2102.04664, 2021.